\documentclass[osajnl,twocolumn,showpacs,superscriptaddress,10pt]{revtex4-1}

\usepackage{amsmath}
\usepackage{amssymb}
\usepackage{amsfonts}
\usepackage{graphicx}
\usepackage{natbib}

\begin{document}

\title{An on-chip optical lattice for cold atom experiments}

\author{Cameron J. E. Straatsma}
\affiliation{JILA and Department of Electrical, Computer, and Energy Engineering, University of Colorado, Boulder, CO 80309-0440}

\author{Megan K. Ivory}
\affiliation{ColdQuanta Inc., 3030 Sterling Circle, Boulder, CO 80301-2338}

\author{Janet Duggan}
\affiliation{ColdQuanta Inc., 3030 Sterling Circle, Boulder, CO 80301-2338}

\author{Jaime Ramirez-Serrano}
\affiliation{ColdQuanta Inc., 3030 Sterling Circle, Boulder, CO 80301-2338}

\author{Dana Z. Anderson}
\affiliation{JILA and Department of Physics, University of Colorado and National Institute of Standards and Technology, Boulder, CO 80309-0440}

\author{Evan A. Salim}
\affiliation{ColdQuanta Inc., 3030 Sterling Circle, Boulder, CO 80301-2338}

\begin{abstract}
An atom-chip-based integrated optical lattice system for cold and ultracold atom applications is presented. The retro-reflection optics necessary for forming the lattice are bonded directly to the atom chip, enabling a compact and robust on-chip optical lattice system. After achieving Bose-Einstein condensation in a magnetic chip trap, we load atoms directly into a vertically oriented 1D optical lattice and demonstrate Landau-Zener tunneling. The atom chip technology presented here can be readily extended to higher dimensional optical lattices.
\end{abstract}


\maketitle

Optical lattice potentials are a versatile tool for the study and manipulation of cold and ultracold gases. They are routinely used in the study of fundamental physical phenomena and quantum simulation~\cite{nphys138,nphys2259}, as well as in applied arenas such as timekeeping~\cite{Hinkley13092013,nat12941} and inertial sensing~\cite{PhysRevLett.75.2638,PhysRevA.54.3165,PhysRevLett.85.4498}. The utility of optical lattices stems from their relative simplicity and the ability to control all aspects of the potential to a high degree of precision. These factors make optical lattice instrumentation a prime candidate for integration with current state-of-the-art compact and transportable quantum gas technology.

Much of the current technology for studying quantum gases is constrained to laboratory experiments due to cost, size, and complexity. However, an ongoing effort in developing compact and transportable Bose-Einstein condensate (BEC) systems~\cite{PhysRevA.70.053606,APL.96.093102,QIP.10.11128,APB.89.00340} has been successful in breaking free of these constraints. At the core of these systems is atom chip technology~\cite{nat413498,PhysRevLett.87.230401}, which enables a degree of integration. For example, on-chip optical detectors~\cite{NJP.12.095005} and cavities~\cite{nat06331} have been demonstrated, as well as the generation of highly configurable trapping potentials through radio-frequency (RF) dressing~\cite{nphys420} and optical projection~\cite{APL.102.084104}.

\begin{figure}[h]
\centering
\includegraphics[scale=1]{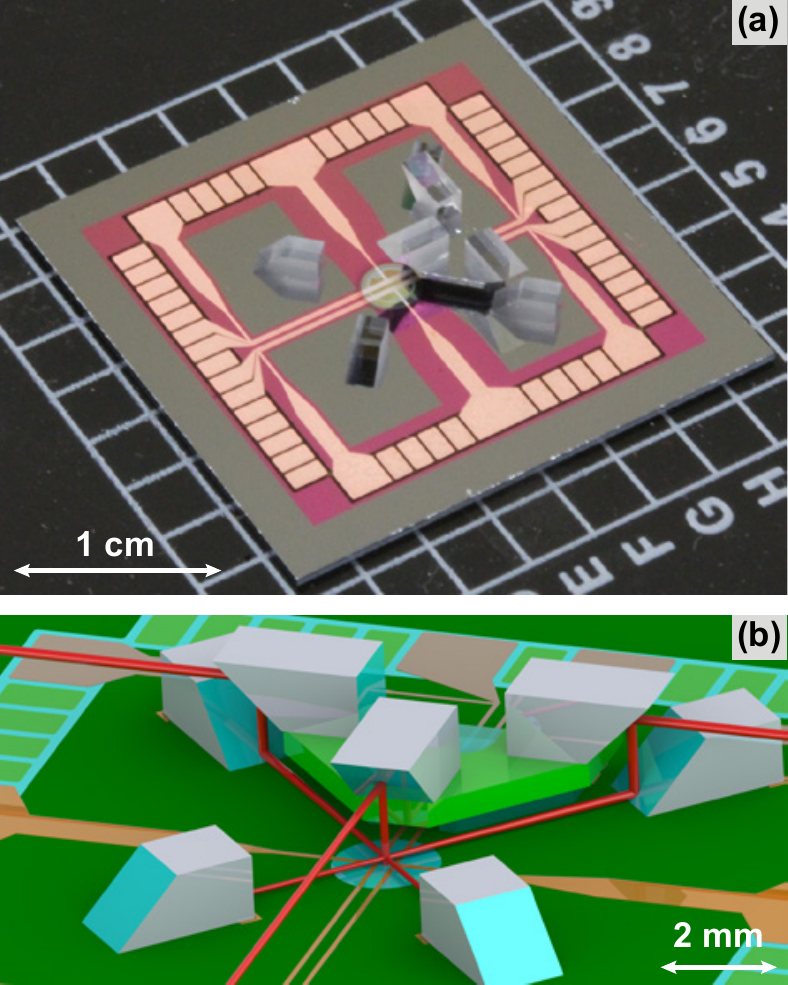}
\caption{\label{fig:chip}(a) Photo of the vacuum side of the atom chip showing the on-chip optical lattice system. (b) Model of the on-chip optical lattice system. The two optics at the bottom of the image retro-reflect two of the incoming lattice beams, while the window is used to retro-reflect the third beam. The periscope optics are shown in the upper right- and left-hand corners of the image.}
\end{figure}

\begin{figure}[h]
\centering
\includegraphics[scale=1]{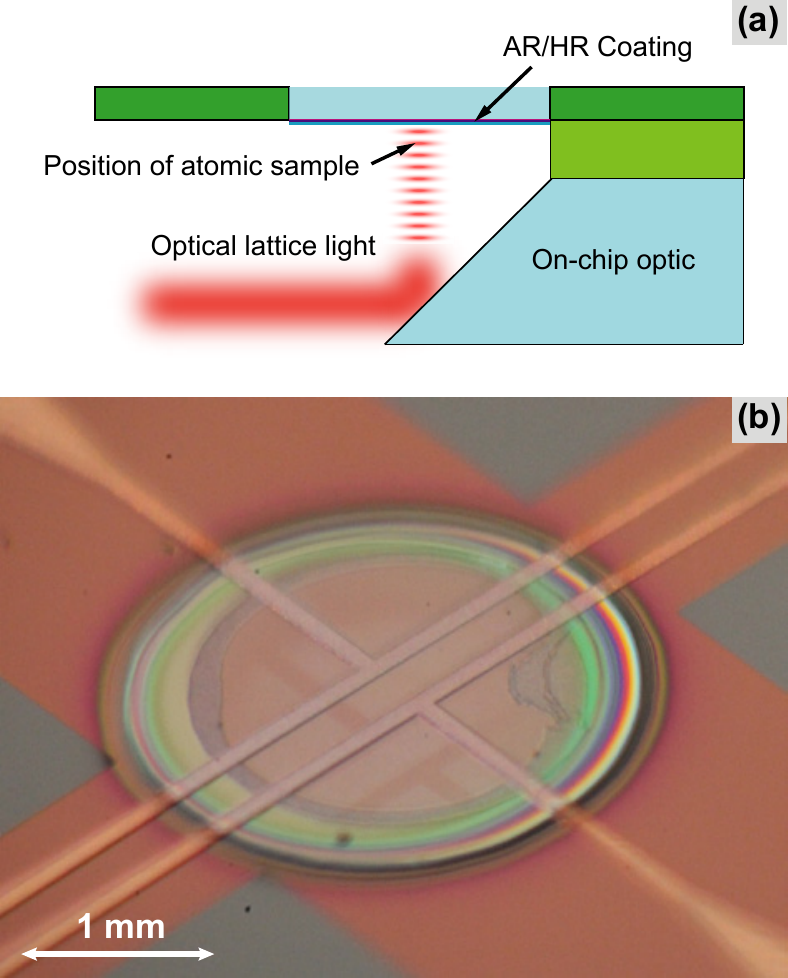}
\caption{\label{fig:lattice}(a) Schematic illustration of the vertically oriented 1D lattice used to demonstrate Landau-Zener tunneling. (b) Photo of the mirror coating on the vacuum side of the atom chip.}
\end{figure}

Gallego et al. were the first to integrate on optical lattice with an atom chip~\cite{OL.34.003463}. To accomplish this, a lattice beam was retro-reflected directly from the surface of the atom chip at almost normal incidence creating a series of pancake-shaped traps parallel to the chip surface. The technology demonstrated in this work is capable of being extended to a fully 3D optical lattice overlapped with a magnetic chip trap. In our case, the magnetic trap is located approximately $100~\mu\text{m}$ from the surface of the atom chip; therefore, the beams necessary for forming a 3D lattice cannot be directly focused onto atoms in the magnetic trap without experiencing diffraction from the edge of the chip. To alleviate the effects of diffraction, miniature optical elements are bonded directly to the atom chip to act as periscopes for each lattice beam. As shown in Fig.~\ref{fig:chip}, in addition to the periscope optics, fixed retro-reflection optics are bonded to the chip for two of the lattice beams. The third lattice beam is retro-reflected from a window embedded in the atom chip substrate. This window also enables through-chip imaging of atoms in the lattice potential at numerical apertures as high as 0.8. Figure~\ref{fig:chip} provides a depiction of the atom chip and the on-chip optics used for generating a lattice potential.
To demonstrate the utility of this system, a proof-of-principle experiment is performed where ultracold atoms are trapped in a vertically oriented 1D lattice, schematically shown in Fig.~\ref{fig:lattice}(a), that holds the atoms against gravity.

Similar to the chips described in~\cite{QIP.10.11128} and~\cite{APL.102.084104}, the atom chip used in this work is a $420~\mu\text{m}$ thick silicon substrate measuring $23\times23~\text{mm}$. The substrate incorporates co-planar regions of glass and silicon that enable high density, through-chip electrical vias as well as in-chip windows, both of which are ultrahigh vacuum compatible. The chip is metalized on both the ambient and vacuum side allowing for the realization of a variety of micro-magnetic atom chip traps. As mentioned, the atom chip used in this work incorporates miniature optical elements anodically bonded directly to the vacuum side of the chip. Notably, the bonding is achieved without use of epoxies that would degrade vacuum quality.

The positioning of the on-chip optical elements is designed to deliver the beams necessary for producing a 3D optical lattice in the vicinity of a magnetic chip trap (see Fig.~\ref{fig:lattice}(a)). It is necessary to be able to align the lattice beams such that they intersect within a small fraction of the beam waist ($w_0\approx100~\mu\text{m}$ here), and that the retro-reflected beams remain collinear over a distance of a few millimeters. To ensure sufficient precision in positioning these optical elements, alignment marks are patterned on the chip surface during fabrication. By inspection with a microscope it is found that the on-chip optics are positioned to within $10~\mu\text{m}$ and $0.1^\circ$ of the alignment marks, which satisfies the precision criteria. In addition to the optics bonded to the chip, there is a $3~\text{mm}$ diameter glass window in the center of the chip. This window is mirror coated (see Fig.~\ref{fig:lattice}(b)) to reflect $>99.9\%$ of light at $1064~\text{nm}$ and $<0.1\%$ of light at $780~\text{nm}$ for $0-10^\circ$ angles of incidence. The dielectric coating has a damage threshold in excess of $10^9~\text{W}/\text{cm}^2$ at $1064~\text{nm}$; therefore, lattice beam powers $>1~\text{W}$ can easily be implemented for beam waists of $100~\mu\text{m}$. The purpose of the window is twofold: $1)$ it allows lattice light to be reflected from the even surface of the chip window rather than from a chip wire, and $2)$ it provides the capability for through-chip, high resolution imaging of atoms trapped in the lattice. Through-chip imaging and optical projection have been demonstrated in~\cite{APL.102.084104}.

The atom chip forms the top of a ColdQuanta RuBECi$^\circledR$ -- a two-chamber, ultrahigh vacuum system designed for the production of rubidium BECs. The RuBECi$^\circledR$ uses a double MOT approach to laser cooling alkali atoms, and has been described in detail elsewhere~\cite{arxiv.1403.4641}. Here, we briefly discuss the differences in the system used for this work, which closely resembles the system described in~\cite{APL.102.084104}. Note that these differences are in addition to the custom atom chip described above.

Similar to the work in~\cite{APL.102.084104}, atoms are loaded onto the atom chip from a compressed 3D MOT using a second pair of quadrupole coils vertically offset and centered at the level of the chip. With this ladder configuration, atoms can be transferred to the chip by ramping down the MOT field while simultaneously ramping on a quadrupole field in the second coil pair. This configuration is used as it leaves the area above the chip open such that a high resolution microscope objective can be placed directly above the atom chip. Although the presence of the on-chip optic for the vertical lattice beam complicates the process of loading the chip trap, on the order of $40\times10^6$ $^{87}\text{Rb}$ atoms remain in the quadrupole trap at the end of the transfer stage. To avoid substantial atom loss, bias fields are used to move the atom cloud around the on-chip optic during the loading process. From the quadrupole trap, $15\times10^6$ atoms are typically captured in the chip trap with trapping frequencies of $642~\text{Hz}$ and $146~\text{Hz}$ along the radial and axial directions, respectively. The atoms are cooled to degeneracy in approximately $1.5~\text{s}$ using forced RF evaporation, which results in quasi-pure condensates with up to $20\times10^3$ atoms. The vertically oriented 1D optical lattice is formed by retro-reflecting a beam with $\lambda = 1064~\text{nm}$ and an estimated $100~\mu\text{m}$ beam waist from the window in the center of the chip.

Following formation of BEC, atoms are loaded directly from the atom chip trap to the optical lattice potential by ramping on the lattice beam to a depth of about $40~E_r$ (where $E_r = h^2/2m\lambda^2$) over $30~\text{ms}$ and quickly extinguishing the chip trap. At this lattice depth the atoms have been held for $>20~\text{ms}$ with no significant heating. Using this method $>90\%$ of the atoms are typically loaded into the optical lattice with a total experiment cycle time of approximately $1.6~\text{s}$. The lattice beam intensity is then rapidly lowered to provide the desired trap depth, held steady for $2~\text{ms}$, and then switched off. The atoms can then be imaged via absorption imaging after a short time-of-flight (TOF) of $3~\text{ms}$.

The periodic structure of an optical lattice leads to a band structure of the energy spectrum for atoms trapped in the lattice. When a BEC is adiabatically loaded into a stationary lattice it occupies the lowest Bloch band around the zero momentum state. However, under the influence of gravity the lattice is tilted and the BEC will begin to accelerate. This acceleration causes the BEC to oscillate within the lowest Bloch band with a period given by the Bloch period, $\tau_B = 2p_r/mg$, where $p_r = h/\lambda$ is the recoil momentum, $m$ is the mass of an atom, and $g\approx9.8~\text{m}/\text{s}^2$ is the acceleration due to gravity. If the lattice is weak (i.e. on the order of the recoil energy) there is a finite probability that the BEC will tunnel into the first excited band when it reaches the edge of the first Brillouin zone. In a weak lattice the first excited band is untrapped resulting in loss of atoms from the lattice. This effect is known as Landau-Zener tunneling, and it can provide an accurate characterization of the optical lattice depth~\cite{PhysRevA.65.063612}.

The probability of an atom tunneling to the first excited band is given by
\begin{equation} \label{eqn:LZT_prob}
P\left(a\right) = \exp{\left[-\frac{\lambda\left(\Delta E\right)^2}{8\hbar^2a}\right]},
\end{equation}
where $\Delta E$ is the band gap energy between the lowest and first excited Bloch bands and $a$ is the acceleration of the atom. Landau-Zener tunneling occurs in the weak lattice regime where the lattice depth is on the order of the recoil energy. In this regime, the band gap energy is well approximated by the lattice depth as long as the cloud density is small enough such that mean-field effects are negligible~\cite{PhysRevA.65.063612}. This is the regime in which the experiment is performed.

\begin{figure}[h]
\centering
\includegraphics[scale=1]{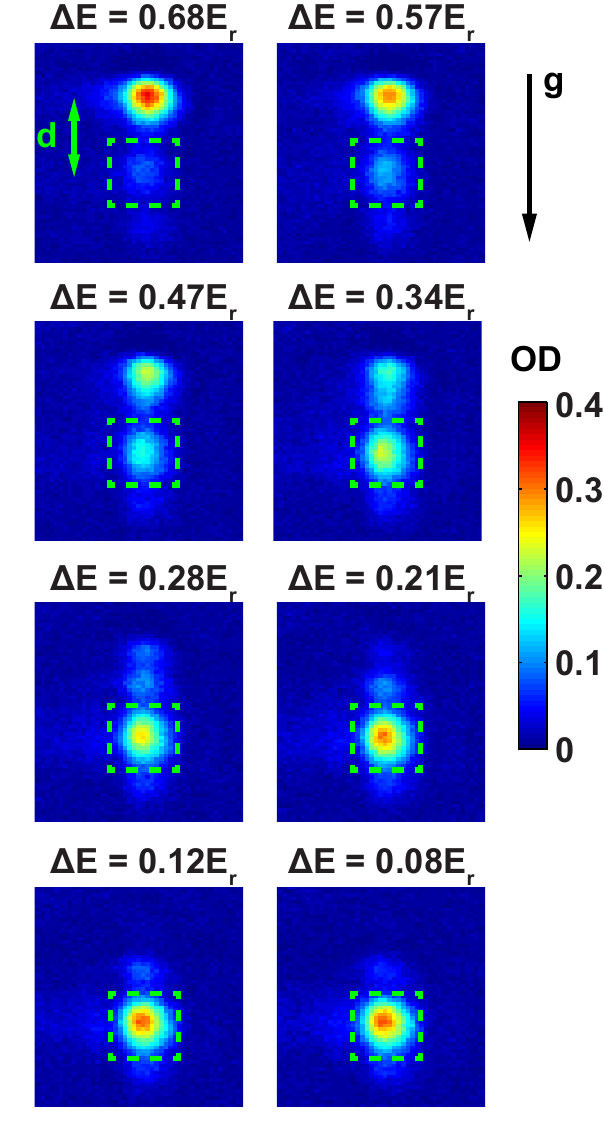}
\caption{\label{fig:LZabs} Absorption images of atoms released from a weak lattice after $3~\text{ms}$ TOF. The boxes (dashed, green) show the region used to calculate the number of atoms that have tunneled out of the lattice, and $d\approx90\mu\text{m}$. The band gap energy quoted is the corrected energy after scaling.}
\end{figure}

In order to verify the presence of the optical lattice and calibrate its depth, a small condensate of approximately $4000$ atoms is formed in the magnetic chip trap and loaded into the lattice. The lattice depth is then ramped to an energy on the order of $E_r$ and held constant for approximately $2.5\tau_B$. This atom number and hold time are chosen such that atoms lost in the first Bloch oscillation are spatially separated from the atoms remaining in the lattice after a short TOF of $3~\text{ms}$. Figure~\ref{fig:LZabs} shows a series of absorption images for different lattice depths.
Clearly evident is the appearance of a large population of atoms separated from those still trapped in the lattice as the lattice depth is decreased.

\begin{figure}[h]
\centering
\includegraphics[scale=1]{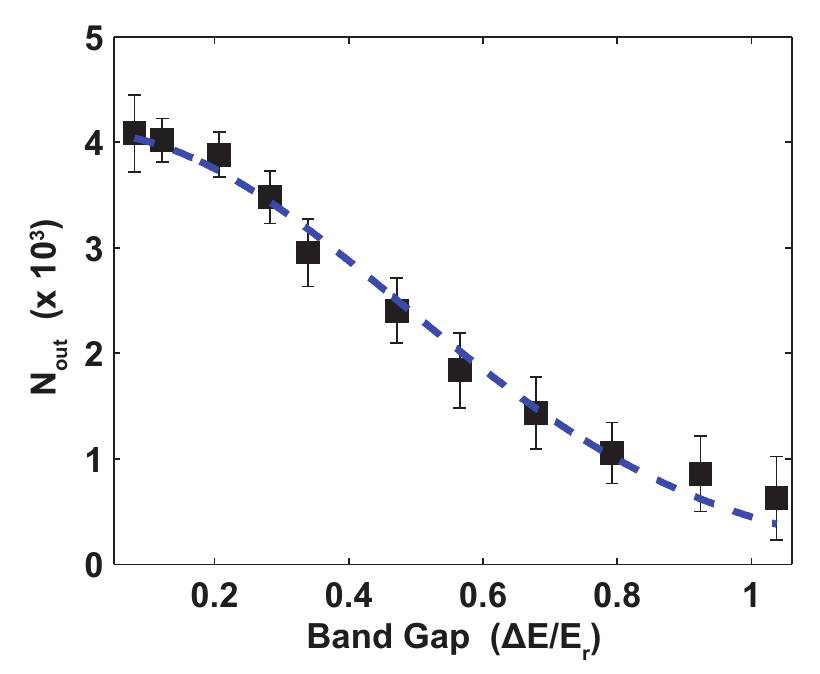}
\caption{\label{fig:LZnum} Number of atoms lost from the lattice in the first Bloch period as a function of the band gap energy. Error bars represent the standard error of the mean of 20 individual realizations of the experiment. The line (dashed, blue) represents a fit of the data to Eqn.~\eqref{eqn:LZT_num}. The band gap energy quoted is the corrected energy after scaling.}
\end{figure}

From the absorption images in Fig.~\ref{fig:LZabs} the number of atoms that have tunneled out of the lattice after the first Bloch period can be calculated and compared to $\Delta E$. The resulting data is shown in Fig.~\ref{fig:LZnum} along with a curve fit based on the number of atoms expected to have tunneled out of the lattice after a single Bloch period:
\begin{equation} \label{eqn:LZT_num}
N_o = N_i\exp{\left[-\frac{\lambda\left(s\Delta E\right)^2}{8\hbar^2g}\right]},
\end{equation} 
where $N_o$ is the number of atoms that tunneled out, $N_i$ is the number of atoms initially in the lattice, and $s$ is a fit parameter for scaling the estimated lattice depth to the actual lattice depth. Note that the band gap energy quoted in Figs.~\ref{fig:LZabs} and~\ref{fig:LZnum} is the actual band gap of the lattice based on the fitted value of $s$. The fit in Fig.~\ref{fig:LZnum} corresponds to $s = 0.75\pm0.05$ implying that the expected depth of the lattice based on input laser power and nominal beam waist size has an error of approximately $25\%$. The reason for this error is due to slight misalignment of the lattice beam and its focusing optics. After optimization it was found that the incident beam was not perfectly centered on the chip window and was clipping a nearby wire on the chip substrate. Additionally, the waist of the beam was not located exactly at the chip window. These factors resulted in smaller lattice depths than expected for a given input laser power.

In conclusion, a method for precisely integrating complex optical structures directly with current atom chip technology has been developed for hybrid magnetic and optical trapping experiments with cold and ultracold atoms. As a demonstration of this technology an on-chip optical lattice system has been constructed and a proof-of-principle experiment with atoms trapped in a vertically oriented 1D optical lattice was performed. In such a configuration the acceleration due to gravity can be used to calibrate the lattice depth based on Landau-Zener tunneling of ultracold atoms from the lowest Bloch band in the lattice.

This work is supported by the U.S. National Science Foundation (NSF) under contract SBIR 1126099 and the U.S. Air Force Office of Scientific Research (AFOSR) under contract STTR FA9550-11-C-0051. The work of C.J.E.S. and D.Z.A. was also partially supported by the U.S. AFOSR under contract FA9550-14-1-0327 and the U.S. NSF under contract PHY1125844.

\end{document}